# Influence of surface pre-deformation on the Portevin-Le Chatelier effect and the related multiscale complexity of plastic flow in an Al-Mg alloy

Hafsa Jaber*, Benoît Beausir, Denis Entemeyer, Tatiana Lebedkina, Marc Novelli, Mikhail Lebyodkin*

Laboratoire d'Etude des Microstructures et de Mécanique des Matériaux (LEM3), CNRS, Université de Lorraine, Arts & Métiers ParisTech, 7 rue Félix Savart, 57070 Metz, France

* Corresponding authors.
E-mail addresses: hafsa.jaber@univ-lorraine.fr (H. Jaber), mikhail.lebedkin@univ-lorraine.fr (M. Lebyodkin)

## Abstract

The influence of the surface pre-deformation on jerky flow caused by the Portevin-Le Chatelier (PLC) effect was investigated using flat tensile specimens of an Al-Mg alloy. Although jerky flow represents a macroscopic plastic instability, the underlying mechanisms stem from self-organization of dislocations, which pertains to deformation processes at mesoscopic scales. To provide a comprehensive approach, the investigation was carried out by coupling tensile tests, digital image correlation and acoustic emission techniques, each targeting a particular range of scales. Thin superficial layers were pre-deformed using surface mechanical attrition technique (SMAT). It was found that the observed effects depend on which surfaces are processed. Overall, the treated samples exhibited an enhanced yield strength without deterioration of ductility in comparison with the initial material. These changes in the general mechanical behavior are likely to be correlated with the changes in jerky flow. Using digital image correlation, a tendency to delocalization of bursts of plastic flow was found for both the PLC bands and smaller-scale strain heterogeneities outside the bands, the latter having been little studied so far. Unexpectedly, SMAT occurred to modify plastic flow drastically at this scale. In addition to clarification of the role played by the surface in the PLC effect, these findings provide new insights into relationships between deformation processes at distinct scales.

*Keywords*: Portevin-Le Chatelier effect; Surface mechanical attrition treatment; Self-organization and patterning; Digital image correlation; Acoustic methods

## 1. Introduction

This paper brings together two research fields, one concerning a long-standing problem of plastic instability in metallic alloys, known as the Portevin-Le Chatelier (PLC) effect [1], and the other related to recent concepts aimed at improving mechanical properties of materials by severe plastic deformation (SPD) of the material surface [2-4]. The interest in the PLC effect is constantly renewed in view of a persistent elaboration of new classes of alloyed materials with different compositions and microstructures [5-9]. The instability occurs as recurrent localizations of plastic deformation giving rise to strain-rate bursts and, consequently, serrated deformation curves, or "jerky flow". Its mechanism is generally ascribed to the negative strain-rate sensitivity of stress (SRS) caused by the dynamical strain aging (DSA), i.e., accumulation of solute atoms on the mobile dislocations temporarily arrested at obstacles [10-13]. The negative SRS stems from a downturn in the solute-induced pinning strength with increasing strain rate because of the respective decrease in the dislocation arrest time. Despite this clear microscopic basis, the prediction of behaviors of real alloys is challenging. Not only real behaviors are often difficult to refer to one of the qualitatively categorized generic types [14], but different authors often come to different conclusions with regard to the influence of various control factors, notably, the alloy composition, the initial microstructure, the strain rate or temperature, on the measured characteristics, such as the critical strain for the instability onset [11,15-17], the amplitude and frequency of serrations [7,16,17], the parameters and kinematics of the deformation bands [18-21]. This difficulty justifies the unceasing interest in the PLC effect. From the practical viewpoint, the objective is to avoid the plastic instability since it leads to undesirable consequences, such as the ductility reduction or surface degradation because of the traces left by the PLC bands [14].

The PLC effect also attracts a growing fundamental interest as a striking example of self-organization in nonlinear dynamical systems. The negative SRS unifies the dislocation ensemble with various systems displaying complex dynamics related to a negative resistance to external force. It was found [9,22] that depending on the experimental conditions, the PLC effect can manifest various collective phenomena, from self-organized criticality (SOC) in extended dynamical systems [23], to chaotic dynamics in low-dimensional nonlinear systems [24], and finally to almost periodic relaxation oscillations attributed to the synchronization phenomenon [25].

SPD methods have attracted a great attention due to their capacity to produce high-strength materials by refining the grain structure down to a nanoscopic range [26,27]. As the gain in the strength of fully nanocrystalline materials goes together with a drastic ductility reduction, the further step involved creation of gradient microstructures. One of the intensely explored ideas is to limit the SPD impact to the surface layers [2-4]. This concept made it possible to improve the strength due to a stronger surface region, keeping at the same time an important ductility level due to the softer core which assures plasticity and postpones the necking. Many approaches of surface treatment were proposed, e.g., shot peening [28], mechanical grinding [29], mechanical attrition [30], and others.

As virtually all industrial materials are alloyed, investigation of the possible influence of the surface strengthening on the PLC effect is of great interest. Therewith, the strain localization being a key element of unstable flow, fundamental information may be expected from combining methods of controlled surface modification and techniques allowing for a survey of the surface strain during mechanical testing. In particular, recent elaboration of high-frequency methods of strain visualization by video-recording, speckle interferometry, or digital image correlation (DIC), allowed to detect various scenarios of the PLC band formation [17,18,21,31-33].

Most attention concerning the surface SPD is currently turned to the search of ways to overcome the strength-ductility trade-off, while very few works examined the PLC effect. An only systematic study was realized in Ref. [34]. It was found that Surface Mechanical Attrition Treatment (SMAT) effectively postponed the onset of the PLC effect in a 5182 Al alloy, thus showing a potential to suppress the instability. At the same time, the amplitude of serrations demonstrated a tendency to an increase. A retardation of the instability was also reported for a gradient microstructure produced by hot compression bonding of several sheets of a 5052 Al alloy [35]. Importantly, all literature data concern samples with a through-thickness gradient microstructure, while the question of the surface effect has not been raised explicitly. Nevertheless, some hints can be found in the literature not deliberately devoted to this problem. For example, a sensitivity of the instability onset to surface modifications has been noticed upon polishing, with different changes produced by a mechanical or chemical finish [36,37].

In the present paper, the surface effect on the PLC instability was studied using SMAT adjusted so as to confine its impact to surface layers, without modifying the entire volume of the sample. Furthermore, in order to reduce additional factors that could influence on the PLC behavior, the study was performed on a model coarse-grained binary Al3%Mg alloy. DIC was applied to visualize the PLC bands and their evolution during tensile tests. Besides, attention was also drawn to finer strain-localization events which may not produce discernible stress fluctuations. This "mesoscopic" scale has been little studied so far in the conditions of the PLC effect. However, the available literature data bear witness that such knowledge may provide valuable information on the development of macroscopic instabilities [9,38]. Moreover, mesoscopic scales present a particular interest from the viewpoint of collective deformation processes even in the absence of macroscopic instabilities [39,40]. These concepts were also supported by recording the acoustic emission (AE) accompanying plastic deformation, which implies a further refinement of the surveyed scale in comparison with DIC [41-43]. This analysis of various material responses to plastic deformation made it possible to examine a multifaceted nature of the surface effect on the heterogeneity of plastic deformation.

## 2. Experimental
### 2.1. Sample preparation and microstructure characterization

Tensile specimens were cut from a cold-rolled sheet of a polycrystalline Al-3%Mg (wt.%) alloy and annealed at 400°C for 2 hours (cf. [43]). The specimens had a dog-bone shape with the gauge length, width and thickness of 35 mm, 7 mm and 2.8–3 mm, respectively. The polycrystalline grains exhibited an approximately equiaxed shape and size ranging between 20 and 160 µm, with an average size about 60 µm.

The SMAT was performed using Ø2 mm 100Cr6 steel shots set in motion by a sonotrode with a vibrating amplitude of 40 µm and a frequency of 20 kHz. The distance between the sonotrode and the sample was set to 40 mm. Several treatment durations, 1 min, 3 min, and 5 min, were tested to determine an optimal regime. According to the results obtained (see Sec. 3.1), specimens for tensile tests were subjected to a one-minute SMAT. The specimens were divided into four groups as specified by the targeted surfaces. Two groups had one or two large surfaces treated (specimens referred to as 1L and 2L, respectively). Narrow side surfaces were treated for the other two groups (accordingly, 1N and 2N). The profile roughness parameter, $R_a$, varied about 0.1 µm before SMAT and increased to 3 µm after the treatment. Two specimens of each type were tested in tensile experiments.

Optical and scanning electron microscopes were used to visualize polycrystalline grains. To assess plastic deformation produced by SMAT in the surface layers, the density of geometrically necessary dislocations (GND) was deduced from crystallographic orientations measured using the Electron Back-Scattered Diffraction (EBSD) implemented on a FEG-SEM Jeol F100 equipped with an Oxford Symmetry camera. The sample-to-detector distance was 15 mm, the accelerating voltage was 15 kV, and the step size was set to 0.45 µm in order to provide a good spatial resolution of the GND repartition. The commercial Aztec software was applied to produce crystallographic orientation maps from the EBSD patterns. The lattice curvature κ determined from the disorientations was then used to obtain GND densities from the Nye tensor, $\boldsymbol{\alpha} \cong tr(\boldsymbol{\kappa}).\boldsymbol{I} - \boldsymbol{\kappa}^T$ [44], which was calculated with the aid of the ATEX software developed in the Lorraine University [45].

Vickers hardness tests were employed to assess the SMAT-induced material hardening throughout the specimen thickness. The tests were conducted using the ZWICK/ROELL Indentec ZHV microhardness tester. A controlled load, $F$ = 50 N, was applied for 10 seconds. Five parallel sequences of indentations were performed on the specimen cross-section along the normal to the treated surface, with a step of 100 µm between indentations and 50 µm from the extreme indentation to the specimen edge. The depth dependence of the Vickers hardness, HV = 0.1891 × ($F / d^2$), where $d$ is the average length of two diagonals of the square trace of the indenter, was determined by taking an average over 5 indentations corresponding to the same depth.

### 2.2. Deformation experiments

Tensile tests were performed using a Zwick 1476 testing machine, with the stiffness of the machine/specimen couple about 6 × $10^6$ N/m. The force noise in the idle state corresponded to a stress level of 0.003 MPa. The specimens were deformed at 20 °C with a crosshead velocity of 0.3 mm/min corresponding to the initial applied strain rate, $\dot{\varepsilon}_a$ = 1.4 × $10^{-4}$ s$^{-1}$. The basic sampling rate of 50 Hz was used to record the force vs. time response, in consistence with the highest recording frequency in DIC measurements (see below). A higher sampling rate of 250 Hz was also applied in time intervals selected to gather a more precise information on the stress fluctuations.

The two-dimensional DIC method is based on recording subsequent images of a speckle pattern deposited on a specimen surface, so that the calculation of the difference between the current and the initial images allows to assess the evolution of the in-plane local strain field during deformation [46]. The setup was similar to that described in [47,48], additionally supplied with a mirror to view a lateral side simultaneously with the large frontal surface. The images of a black-and-white speckle patterns were captured by a Charge-Coupled Device camera which allowed for a recoding frequency up to 50 fps with a spatial resolution better than 25 µm/pixel. As the test duration attained 2000 s and longer, such an acquisition rate was only used for selected time intervals. Otherwise, the frequency was reduced to 10 fps. The fields of local displacements in tensile direction and the corresponding strain and strain-rate fields were calculated with the aid of the standard Vic-2D software [49], using subsets of 21 pixels and a step size of 5 pixels.

As the investigations did not only aim at the macroscopic PLC instability but also at smaller-scale plasticity events, special precautions were taken to avoid detecting false strain-rate fluctuations that may be caused by the instrumental noise, e.g., vibrations of the deformation machine and camera or lighting fluctuations. For this, only the strain-rate bursts exceeding the maximum level of the signal recorded before the beginning of loading were considered to be true. As the maximum level included all outliers, this strict choice of the threshold inevitably led to a loss of useful information but assured correct qualitative conclusions.

Besides, AE was recorded simultaneously with the deformation curve and speckle patterns. In order to have a basis for comparison with the literature data for AlMg-based alloys, the experimental setup and data processing adapted the approach described in [43,50]. The AE was captured by a piezoelectric sensor with a wide frequency band of 200 to 900 kHz, clamped to the sample head just above the gauge length. The sonic contact was controlled with the aid of the Nilsen test [51]. The signal from the sensor was pre-amplified by 40 dB and recorded by a Euro Physical Acoustic system with a sampling frequency of 2 MHz [52]. The AE events (hits) were extracted in the real time using several preset parameters, namely, an amplitude threshold of 25 dB, a hit definition time, HDT = 50 µs, and a hit lockout time, HLT = 100 µs. In this scheme, an event is considered to start when the signal exceeds the threshold and to terminate at the time instant after which the signal remains under the threshold for at least HDT. The HLT is conventionally provided to filter out secondary acoustic echoes. The HLT means a dead time during which no records are made since the last recorded hit. Finally, a peak definition time, PDT = 40 µs, is used to detect the peak amplitude of the hit. The so-recorded series of events allow for a statistical analysis of the AE during the entire tensile tests. The acoustic system also allows to store continuous AE records which provide access to waveforms of individual AE hits (see, e.g., [53,54]). This aspect however goes beyond the scope of the present work and will be published elsewhere.

## 3. Results
### 3.1. Vickers hardness and microstructure analysis

Fig. 1 presents examples of measurement of the material hardness as a function of the distance from the surface subjected to SMAT. The initial hardness value varied about the level of 45 kgf/mm$^2$ for all tested samples (cf. curves *1*). The example presented in Fig. 1(a) shows an increase to approximately 73 kgf/mm² near the impacted surface immediately after one-minute SMAT (curve *2*). Further relaxation led to a slight decrease in this extreme value which stabilized at a level of approximately 67 kgf/mm² (curve *3*). The illustrations provided in Fig. 1 also demonstrate that the hardening effect rapidly decays with the distance from the treated surface and fades at the depth of about 0.5 mm, leaving intact a significant part of the bulk. Finally, curve *2* in Fig. 1(b) shows that the difference in the hardness between the treated surface and the volume was considerably levelled off after tensile deformation.

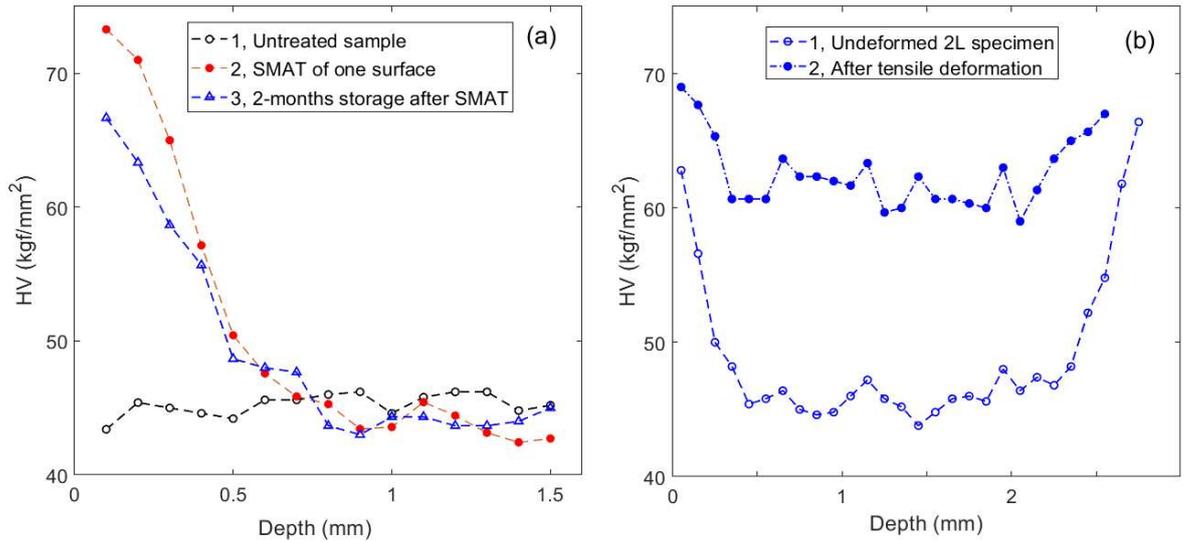

*Figure 1.* SMAT effect on the surface hardness. (a) Dependence of the Vickers hardness, HV, on the depth under the surface treated by SMAT: 1 – Untreated specimen; 2 – immediately after one-minute SMAT; 3 - after two-months storage following the SMAT. (b) Example of changes induced by tensile deformation: 1 – 2L specimen before tensile test and 2 – after tensile test (see Fig. 3).

Qualitatively similar results were obtained for the treatment durations of 3 and 5 minutes, with a slightly stronger hardening effect. Importantly, the increase in the SMAT duration resulted in deeper modifications, increasing the thickness of the transitory layer to 1 mm and more (cf. [34]). To leave intact a large part of the material bulk, one-minute SMAT regime was chosen for the present study. An additional reason for this choice is that one-minute treatment involves a thickness typically affected in the processes of specimen preparation, e.g., mechanical polishing or electrical discharge cutting. The tensile tests were performed after storing the treated specimens for two months. Therewith, tensile tests on the unprocessed material verified that the natural ageing during this period did not produce any change in either mechanical properties or the PLC instability.

The results of the hardness tests were confirmed and refined by the comparison of GND maps after various treatments. Fig. 2 illustrates such results for specimens subjected to SMAT for 5 minutes (a) and 1 minute (b). It can be seen that SMAT induced distortions in a surface layer, visualized by the spatial distributions of the GND density. In order to quantify the depth affected by the SMAT, the average GND density was calculated along the normal to the processed surface. The so-obtained evolution in the depth is plotted in Fig. 2(c). The SMAT effect decreased with decreasing the treatment duration, as evidenced by both the reduced thickness of the impacted layer and lower GND densities ($1.2 \times 10^{14}$ and $4 \times 10^{13}$ $m^{-2}$ for 5 and 1 minutes, respectively). The most significant distortions are concentrated in a layer

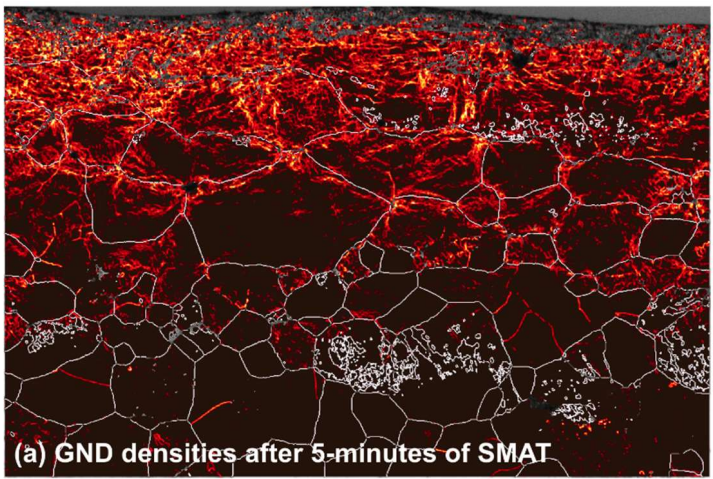

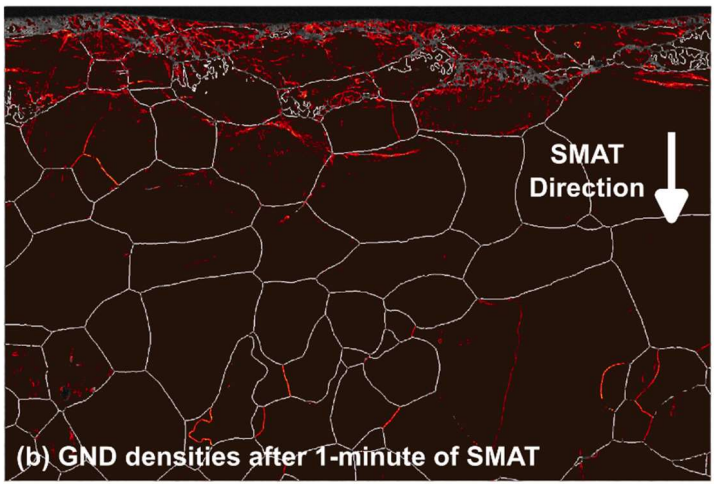

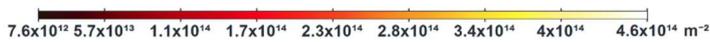

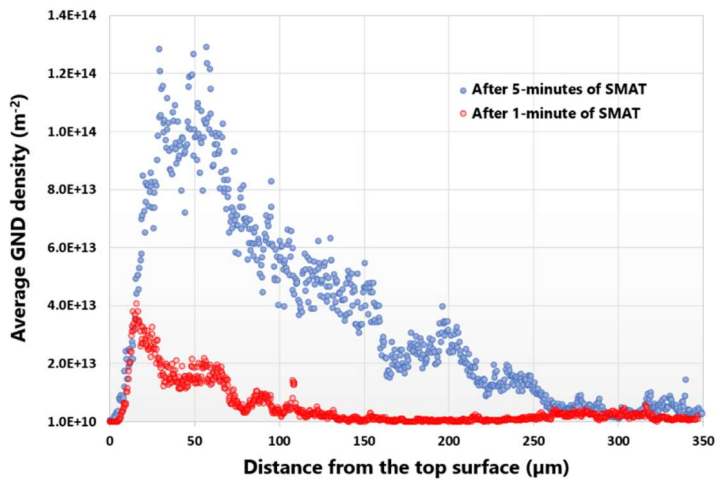

(c) Evolution of GND densities from the SMAT surfaces

approximately 150 µm thick for the 5-minutes SMAT and less than 50 µm thick after 1-minute treatment. Nevertheless, dislocations are still visible in deeper layers, in agreement with the gradual decrease in the Vickers hardness to the depth of 0.5 mm or beyond, depending on the SMAT duration. At the same time, it is noteworthy that although the high GND density attests to a strong plastic deformation near the treated surface, Fig. 2 reveals a similar grain size in the surface region and in the bulk.

*Figure 2.* Geometrically necessary dislocation (GND) densities after (a) 5 minutes and (b) 1 minute of SMAT. (c) In-depth evolution of average GND densities.

### 3.2. Overall mechanical behavior

Fig. 3 illustrates the effect of different SMAT combinations on deformation behavior of the investigated alloy. The examples of entire stress-time dependences, $\sigma(t)$, presented in Fig. 3(a), demonstrate that the treated samples exhibit an overall higher deforming stress than the initial material.

The excess of stress is maximum upon the elastoplastic transition and diminishes progressively towards the onset of necking, where the curves converge and show a similar ultimate tensile strength. Summarizing all tests, the impact may be ranked in the following order according to its magnitude: 1N < 2N ≲ 1L < 2L. For convenience, 1N, 2N and 1L regimes will be referred to as "medium treatments". The magnification of the initial portions of the engineering deformation curves, $\sigma(\varepsilon)$, illustrates that for 2L samples, the offset yield point is increased by about 40% at the conventional 0.2% strain level and about 50% around 1% of plastic strain [Fig. 3(b)]. The latter number is more representative as it corresponds to work-hardening behavior stabilized after the elastoplastic transition, which course visibly differs between the initial and processed materials.

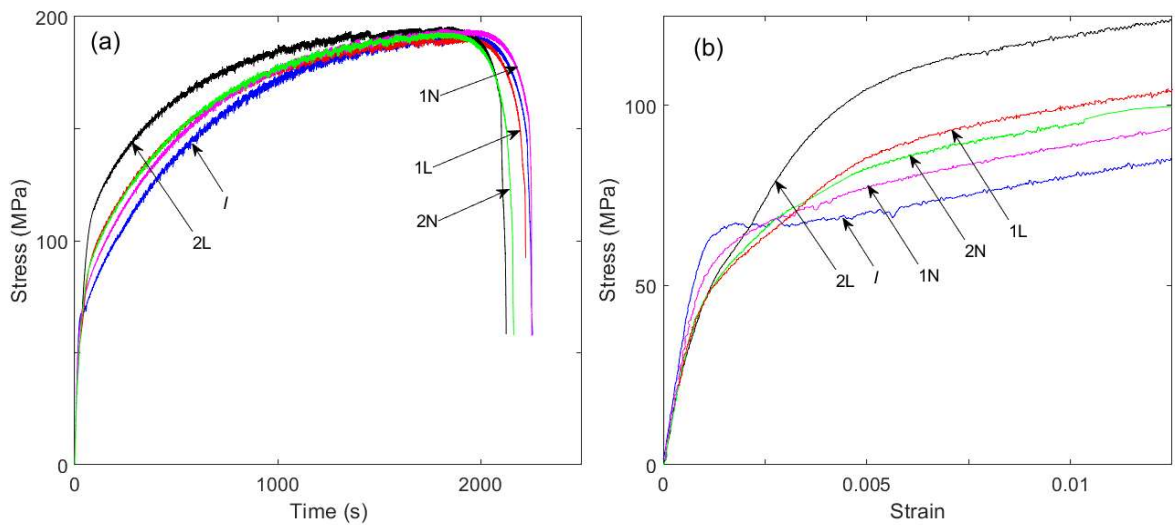

*Figure 3.* SMAT effect on tensile behavior. (a) Examples of stress-time curves for the initial material, designated as I, and specimens subjected to SMAT of various surfaces: 1L and 2L refer to SMAT of 1 and 2 large sides, respectively; 1N and 2N are similar designations for the narrow sides. (b) Magnification of the corresponding engineering deformation curves displaying details of the elastoplastic transition, including the Lüders plateau for the untreated material and the onset of the PLC instability.

Despite the increase in the material strength, Fig. 3(a) testifies that the treated samples do not undergo a significant loss in ductility in comparison with the initial material. More exactly, the variations observed between differently processed samples did not exceed the variations within pairs of samples of the same kind. The persisting ductility is likely to be related to the convergence of the deformation curves. Both these observations are also conforming to the approximately uniform Vickers hardness throughout the specimen thickness after tensile deformation of SMAT samples.

The above-mentioned influence of SMAT on the elastoplastic transition is pertinent to the problem of the strength-ductility interplay. It can be noticed in Fig. 3(b) that the initial material displays a Lüders plateau that suggests propagation of a localized deformation band along the specimen. The process takes place at a non-increasing stress level because the band advances towards unhardened material [55]. The further deformation displays a usual work-hardening behavior concomitant with the PLC effect [43]. It occurs that SMAT of any surface makes disappear the Lüders effect. Two transformations can be acknowledged. First, in contrast to the yield point, the proportionality limit becomes reduced, i.e., the detectable plasticity occurs earlier in SMAT specimens. This effect follows the same ranking regarding the SMAT regime as specified for the yield point, with the maximum reduction of about 50% after 2L treatment. However, soon afterwards the deformation curves of SMAT specimens demonstrate an up-turn that reflects a jump in the work-hardening rate. This up-turn, obviously contributing to the increase in the deforming stress after SMAT, is commonly reported for materials with gradient microstructures and is considered at the heart of their capacity to maintain a good ductility simultaneously with the enhanced strength [35,56].

An insight into the concomitant deformation processes was provided by DIC observations. The propagation of a Lüders band for the initial specimen from Fig. 3 is visualized in Fig. 4 by displaying several consecutive images of the local strain rate field, $\dot{\varepsilon}_{loc}(x, y)$. The numbers above the images reveal that the maximum $\dot{\varepsilon}_{loc}$ inside the band is more than 10 times higher than $\dot{\varepsilon}_a = 1.4 \times 10^{-4}$ s$^{-1}$. The band moves along the specimen (y-direction) at a speed of about 1 mm/s. Consecutive series of such DIC images also allow to visualize the evolution of the propagation pattern during long periods. For this, a one-dimensional (1D) local strain-rate field measured along a selected line segment is traced at consequent time instants. Fig. 5(a) presents a so-obtained "spatiotemporal map" which depicts the function $\dot{\varepsilon}_{loc}(y, t)$ for the tensile axis of the same specimen. For convenience, the dashed-and-dotted lines corresponding to the instant of 48.6 s clarify the correspondence between Figs. 4 and 5(a). In such maps, the propagation of a deformation band appears as an inclined stripe of enhanced color, with its slope and vertical cross-section providing estimates of the propagation velocity and the band width, respectively. It is noteworthy that y-coordinate stands for the relative positions along the specimen gauge with respect to the initial frame. This common choice makes convenient the qualitative comparison of the maps (cf. an alternative approach in [57]), but the specimen elongation should be born in mind for quantitative calculations. It can be seen that one Lüders band, indicated by an arrow, was nucleated near the middle of the gauge length and moved down to the bottom end of the specimen. Then another band occurred above the part hardened by the first band and moved upwards. In addition to the propagation pattern, Figs. 4 and 5 clearly reveal a nonuniform and nonstationary nature of Lüders bands. To illustrate the variations in the band shape and local strain rate, two middle images in Fig. 4 represent subsequent time instants separated by 0.1 s. Fig. 5 also highlights correlations between these variations and stress fluctuations. Finally, Fig. 4 testifies that the strain-rate field outside the Lüders band is only relatively uniform and contains alternating loading and unloading regions corresponding to $\dot{\varepsilon}_{loc}$ variations with lower intensity than within the Lüders band. These fluctuations appear as a dispersion of warmer and colder spots on the maps of Fig. 5. They will be inspected in Section 3.4.

The left-side parts of Figs. 5(b), (c) and (d) illustrate the evolution of the local strain-rate field around the instants corresponding to stress up-turns (see arrows) during deformation of SMAT samples. The patterns occurred to be dependent on the SMAT regime. Fig. 5(b) demonstrates that the strain-rate field was highly uniform before the onset of the PLC effect in 2L samples, and no noticeable change was detected upon the stress deflection. On the contrary, the samples subjected to other treatments displayed a much stronger initial heterogeneity that showed up as a relatively high overall density and/or intensity of the above-mentioned spots. Examination of different patterns revealed a tendency for these heterogeneities to become stronger upon the stress up-turns [see, e.g., Fig. 5(c)]. Fig. 5(d) illustrates a particular behavior observed for one of the 1L samples. In this case, the deflection was accompanied by the occurrence of mobile deformation bands with a low intensity, a kind of precursors of the further PLC bands, which moved with the velocity similar to that of the PLC bands but ran over shorter distances and did not produce noticeable stress fluctuations.

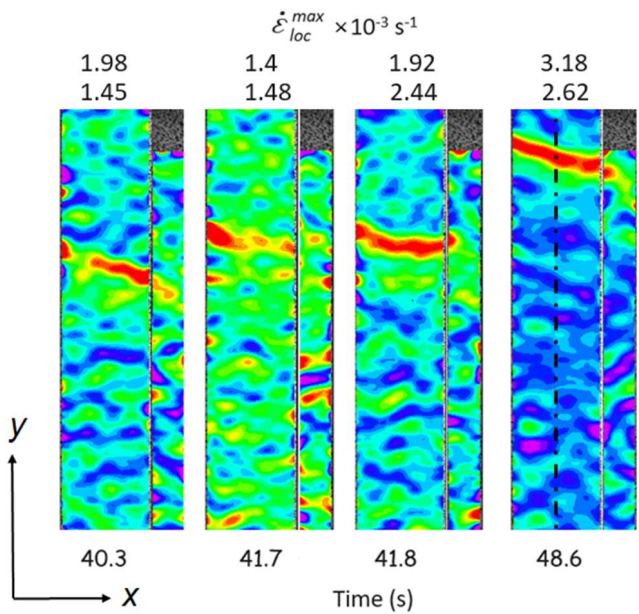

*Figure 4.* Example of Lüders band behavior in an initial sample. The band kinematics is illustrated using consecutive DIC images of the local strain rate field, $\dot{\varepsilon}_{loc}(x, y)$, calculated from speckle photographs recorded at a frequency of 10 fps. The initial height of the gauge part is 34 mm. The vertical centerline depicted by dashes and dots on the image at 48.6 s explains the construction of the maps of Fig. 5 by collecting data along a selected line segment from multiple DIC images. To provide a good contrast for each pattern, the color scale range was not selected once for all frames but adapted to each image. For the sake of comparison, the numbers above the images indicate the maximum $\dot{\varepsilon}_{loc}$ inside the band traces. The upper row corresponds to the large surface and the second row provides data for the narrow surface.

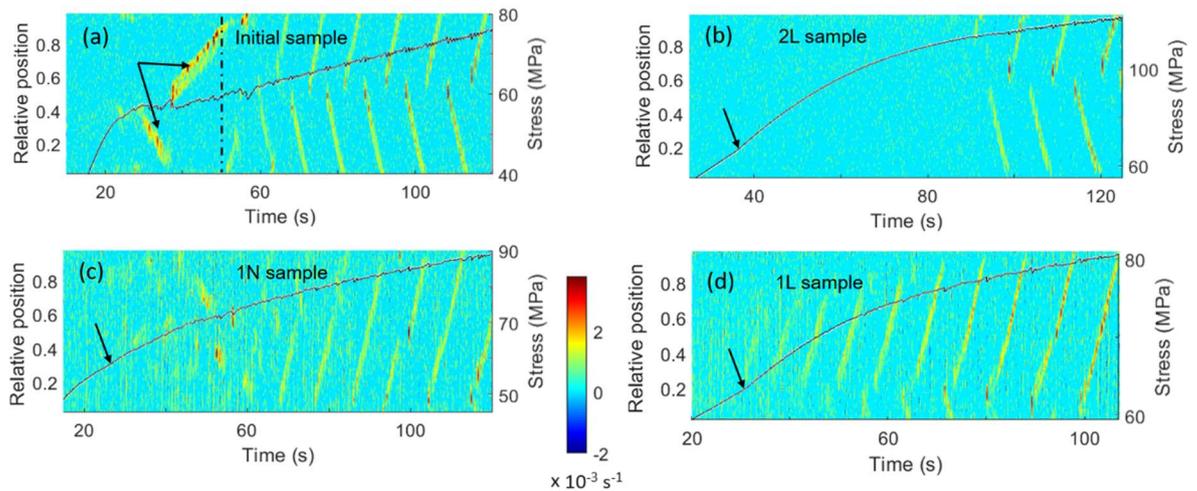

*Figure 5.* Spatiotemporal maps sketching the evolution of the local strain-rate along the tensile axis of (a) an initial specimen, (b) 2L sample, (c) 1N sample, and (d) 1L sample. The time-stress curves are superposed on the maps for convenience. The dashed-dotted line in Fig. 5(a) corresponds to its counterpart in Fig. 4. The y-positions are shown in the range from 0.02 to 0.98 in order to cut off border effects biasing the results of calculations. The arrows indicate Lüders bands in Fig. 5(a) and stress up-turns in the other charts.

Two more qualitative observations are noteworthy. Some of the local strain-rate heterogeneities could be strong enough to produce observable stress fluctuations, thus making ambiguous the conventional detection of the onset of the PLC instability on the deformation curves, as described in Section 3.3 (cf. [58]). In particular, Fig. 5(c) illustrates formation of rather strong short-term deformation bands during the elastoplastic transition in a 1N sample, which may be associated with stress drops (see an interval around 40 s). This feature was especially pronounced for 1N and 2N samples. On the other hand, only some of stress fluctuations could be put into correspondence with well-defined local strain-rate bursts, while others were caused by heterogeneities distributed over the specimen gauge.

**3.3. Influence on the PLC instability**

Besides overall mechanical behavior, it can be seen in Fig. 3 that all specimens demonstrated stress serrations caused by the PLC effect. Fig. 6 describes the influence of SMAT on the critical strain and stress values determined in two ways. Namely, (1) $\varepsilon_{cr1}$ and $\sigma_{cr1}$ correspond to the occurrence of the first sporadic serrations observable on the macroscopic scale of Fig. 5; (2) $\varepsilon_{cr}$ and $\sigma_{cr}$ point to the onset of the stabilized PLC instability (see Fig. 7(a) for detail). The data on $\varepsilon_{cr1}$ and $\sigma_{cr1}$ bear witness that SMAT delays the approach of the instability. Moreover, Fig. 6(b) demonstrates that despite the different nature of the elastoplastic transition in the initial and processed materials, $\sigma_{cr}$ follows a trend similar to that for $\sigma_{cr1}$. The applied stress can thus be considered as a relevant control parameter pertaining to the instability conditions. Therewith, the closeness of $\sigma_{cr}$ and $\sigma_{cr1}$ for 2L samples confirms that this treatment effectively suppresses sporadic instabilities and maintains uniform plastic flow until a major part of the material will approach the critical conditions [cf. Fig. 5(b)]. On the other hand, even if the initial material shows PLC serrations immediately after the Lüders plateau, the strain accumulated due to the passage of the Lüders bands results in higher $\varepsilon_{cr}$ values than those obtained after medium treatments. Accordingly, $\varepsilon_{cr}$ does not demonstrate the same ranking for the SMAT effect as described above but manifests a minimum for 2N samples. Nevertheless, the stronger delay of the instability after 2L treatment results in the highest $\varepsilon_{cr}$ values.

The qualitative evolution of serrations and PLC bands is illustrated in Fig. 7. The behavior was similar for all specimens and displayed types usually reported for various Al alloys at similar strain rate $\dot{\varepsilon}_a$ (see, e.g., [14]). First, type *A* pattern sets on at $\varepsilon_{cr}$, as can be recognized on the left-hand side of Fig. 7(a). This type is characterized by stress rises needed to nucleate new PLC bands, which are followed by stress drops at the moments of nucleation and then a relatively smooth plastic flow during a quasi-continuous propagation of the band along the tensile axis. The propagation velocity was similar for different samples and varied about 4 mm/s. At further deformation, the plastic flow associated with the band propagation became less smooth and displayed serrations related to strain-rate fluctuations within the propagating band [right-hand side of Fig. 7(a)]. This transitory behavior is usually referred to as type *A+B*. Fig. 7(b) illustrates the further transition to a pure type *B*. Therewith, the quasi-continuous propagation changes to the so-called "hopping" or "relay-race" propagation, i.e., the occurrence of correlated sequences of localized PLC bands. Each band gives rise to a stress drop, so that their series are associated with more or less regular serrations resembling relaxation oscillations [59]. Like in the case of type *A*, the beginning of each sequence is often marked by a stress rise followed by a deep drop, which correspond to the nucleation of a PLC band in the material strain-hardened by the passage of the previous group of bands. Fig. 1(a) testifies that this behavior persisted until fracture. The band propagation velocity can also be determined for the hopping propagation. Its evolution was also similar for different samples and corresponded to a general observation that the band propagation slows down with strain [33,47,60]. As a matter of example for the studied alloy, the respective velocity decreased to a value of about 1 mm/s around ε = 0.12. As far as the influence of the surface state is concerned, an effect of SMAT on the described patterns was detected on the initial stage of jerky flow, as a tendency to a faster transition to type *B* patterns of serrations and PLC bands, most clearly identified for 2L specimens.

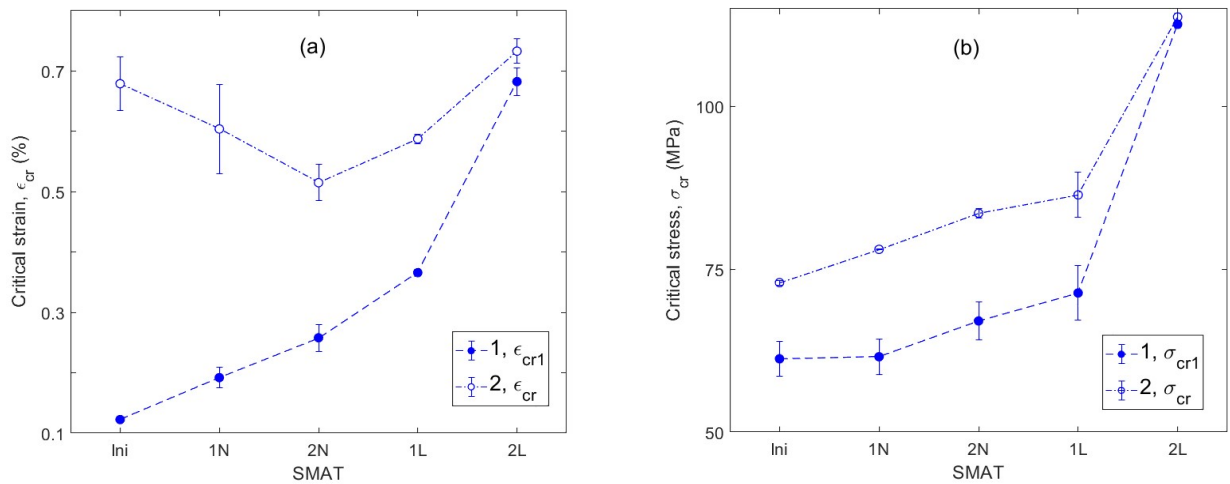

**Figure 6.** *SMAT effect on the (a) critical strain and (b) critical stress for the onset of plastic instability. 1 – $\varepsilon_{cr}$ and $\sigma_{cr1}$ correspond to the occurrence of the first sporadic serrations; 2 – $\varepsilon_{cr}$ and $\sigma_{cr}$ correspond to the onset of stabilized serration patterns. Data points are connected by dashed lines for convenience. Error bars in this and further figures correspond to the variation between values for each pair of samples of the same kind.*

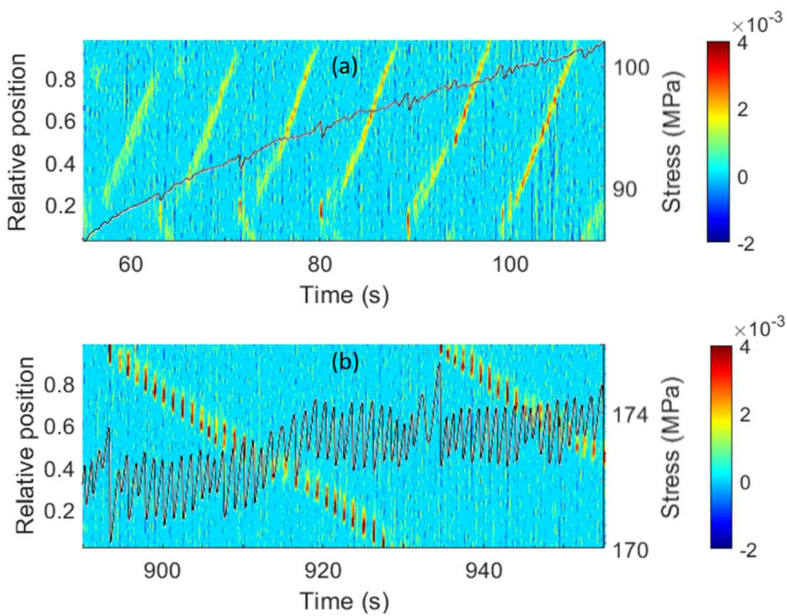

**Figure 7.** *Illustration of (a) the initial type A and type A+B behaviors and (b) stabilized type B pattern for a 1L specimen.*

Besides the spatiotemporal maps, sequences of 2D DIC images were examined to assess the band nucleation and development across the specimen at different stages of deformation. Two major scenarios of band emergence were found for the investigated material. The first three frames of Fig. 8(a) illustrate the well-known scenario of the band nucleation on a lateral side and quick growth toward the opposite side [18,31,32]. Although the image acquisition frequency was relatively low for a detailed survey, analysis of many similar events allowed to unveil some details of such processes. It occurs that this scenario may enclose a variety of patterns, including an incomplete growth followed by the band shrinking, a quasi-instantaneous, in terms of the time resolution of 20 ms, rotation of the band orientation, band splitting, and so on. For example, Fig. 8(a) reveals a peculiar situation when the progressing deformation band triggers a counter band on the opposite side of the sample (frame 2). Fig. 8(b) represents an alternative scenario, the possibility of which was reported recently [21]. It consists of the formation of a well-delineated PLC band (frame 3) from a diffuse deformation band which occurs simultaneously through the entire cross-section of the specimen (frames 1 and 2). Overall, the first scenario of the nucleation and growth of a band nucleus was mainly observed at early stages of plastic flow in the initial material and after SMAT of narrow sides. The simultaneous formation via diffuse bands was found at all deformation stages and was predominant at larger strains. Therewith, this scenario was preponderant at all strains in samples with large sides treated by SMAT. Finally, the band disappearance followed similar schemes, varying between the band shrinking toward one of the lateral sides and the diffuse "dissolution". The latter mode can be recognized in both examples presented in Fig. 8.

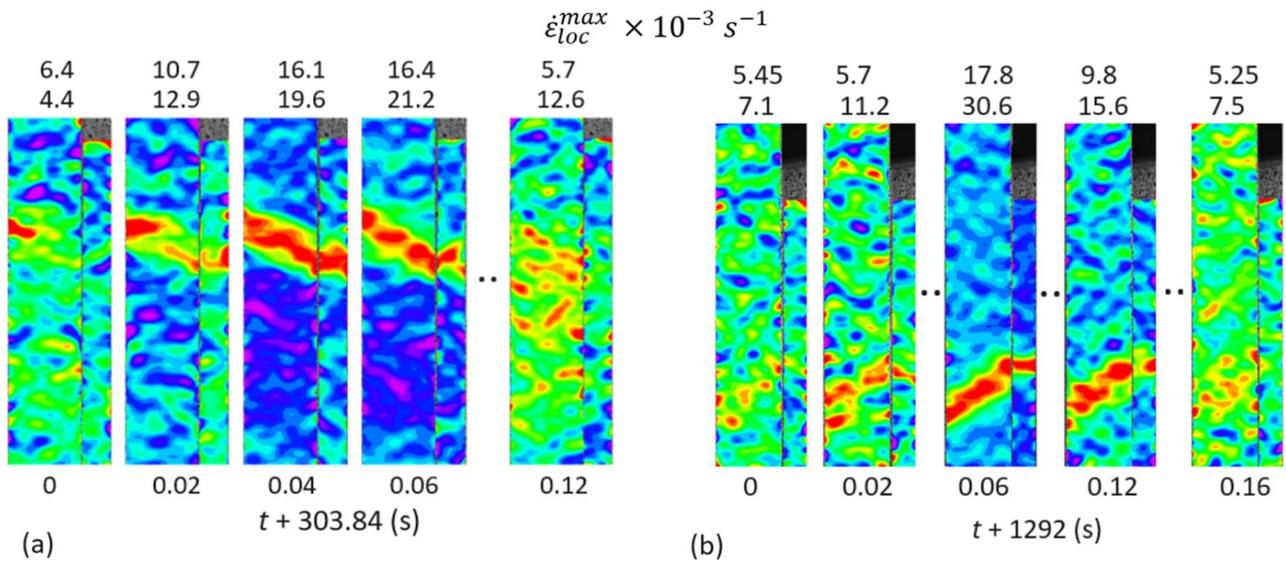

***Figure 8.*** *Series of DIC images illustrating nucleation and formation of PLC bands. (a) 1N sample, early stage of serrated flow; (b) Initial sample, stabilized serrated flow. The upper row of numbers above the images corresponds to the large surface and the second row provides data for the narrow side (cf. Fig. 4).*

The observation of the convergence of the deformation curves (Fig. 3), of similar patterns of serrations (Fig. 7) and similar PLC band kinematics (Fig. 8) established after some deformation for all kinds of specimens could lead to a suggestion that their mechanical behavior is only sensitive to SMAT before the bulk becomes sufficiently work-hardened. Unexpectedly, the statistical analysis of stress serrations in the strain interval $\varepsilon \approx [0.1; 0.15]$, which corresponds to virtually stabilized amplitude of serrations for all samples, unveiled an influence of SMAT on the intensity of plastic instability even after significant work-hardening. One can easily anticipate from Fig. 7(b) that the amplitudes of the stress drops associated with the PLC bands, $\Delta\sigma$, are described by peaked distributions around an average value, as illustrated in the inset of Fig. 9(a). The dependence of $<\Delta\sigma>$ on SMAT indicates that the medium treatments noticeably reduced the serrations (data *1*). 2L treatment almost resumed the serration magnitudes, so that specimens showed $<\Delta\sigma>$ values approaching the initial level. At the same time, the average frequency of serrations, *f*, estimated over the same interval, followed a reverse dependence (data *2*), so that the total sum of $\Delta\sigma$ only varied by about 10% between different samples. That is, the suppression of the intensity of the instability was partly compensated by its increasing "activity", and the contribution of the unstable plastic flow to the total plasticity remained similar for all specimens.

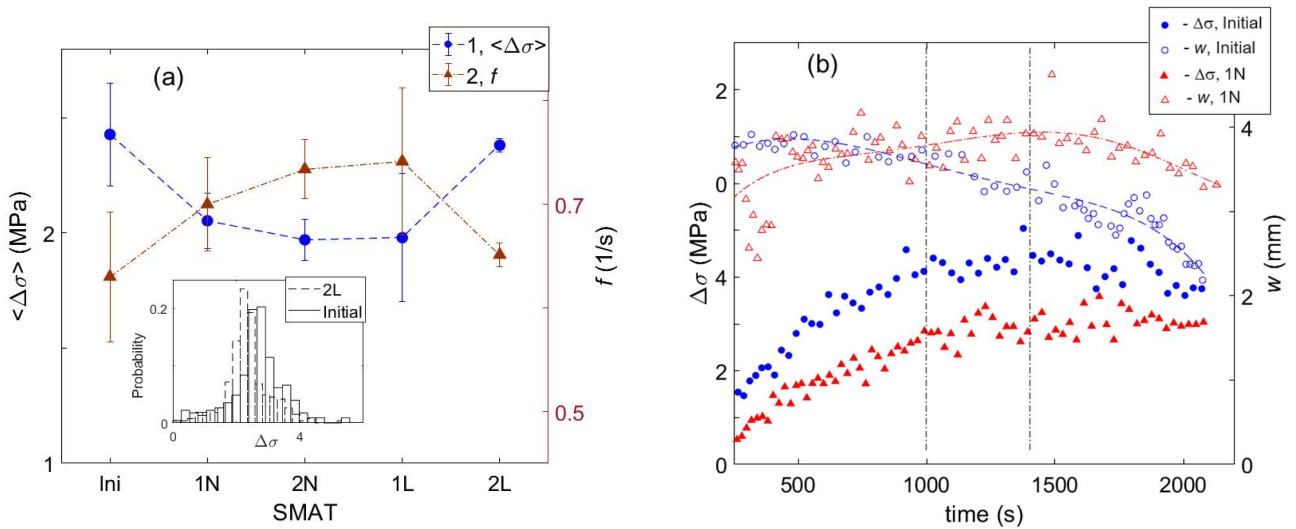

*Figure 9. SMAT effect on various parameters of the PLC instability. (a) 1 – Average amplitude of stress drops, $<\Delta\sigma>$, and 2 – average frequency, f. The averaging was performed over the interval t = [1000 s; 1400s], or $\varepsilon \approx [0.1; 0.15]$, corresponding to virtually stabilized serration amplitudes. The end of the interval was selected far enough from the onset of necking. Inset: Examples of histograms of $\Delta\sigma$ distribution for an initial and 2L samples, traced by a solid and a dashed line, respectively. (b) Evolution of $\Delta\sigma$ (solid symbols) and of the PLC band width, w (open symbols), for an initial (circles) and 1N (triangles) samples. Each data point was obtained by averaging over 20 consecutive instability events. The vertical dashed lines depict the time interval used to evaluate $<\Delta\sigma>$ and f in Fig. 9(a).*

The simultaneous fall in $<\Delta\sigma>$ and increment in *f* lead to a suggestion of a delocalization of the plastic instability after medium treatments. This conjecture is corroborated by Fig. 9(b) which presents an example of comparison of the evolution of the PLC band width, *w*, and the corresponding stress drop amplitudes for an initial and 1N samples. It can be recognized that the initial material displays a descending *w*(*t*) dependence after some initial growth. In contrast, whereas the bands observed in the treated specimen are narrower at small strains, they show a growing tendency until the onset of necking and become considerably wider than in the initial specimen in the analyzed time interval. This tendency was also discerned for the 2L case, although it was less pronounced than after medium SMAT. Therewith, the comparison of the maximum $\dot{\varepsilon}_{loc}$ values brought evidence that the Δσ depression after SMAT was caused by a decrease in the local strain rate within the bands.

Interestingly, the conjecture of delocalization is conforming to the control measurements of the surface roughness which were performed on the untreated large surfaces after tensile tests. Indeed, the SMAT specimens displayed smoother roughness profiles. As a matter of example, the parameter $R_a$ grew to approximately 2.6 μm after deformation of the initial material (cf. Sec. 2.1), but its value varied about 2.0 μm for a 1L and 1N samples.

Furthermore, calculation of higher-order moments revealed modifications in the shape of Δσ distributions after SMAT. For this purpose, the skewness, which characterizes the distribution asymmetry, and kurtosis, which reflects the degree of deviation of the outliers from the mean, were evaluated [61]. Although a comparison between the effects of different treatments was impeded by rather strong variations in the histograms, all SMAT specimens showed a similar qualitative trend regarding the initial state. Like for many characteristics discussed above, except for the average <Δσ> and *f* presented in the previous paragraph, the maximum effect was found for 2L samples. Namely, the kurtosis shifted from the initial value of about 4 to approximately 6 in the 2L state. Whereas the former is comparable to the value of 3, characteristic of the normal distribution, the latter testifies that SMAT leads to essentially stronger deviations from the mean amplitude. The concomitant increase in the skewness, which moved from negative to positive values, suggests an increase in the relative weight of the large-scale tails of the distributions, i.e., higher probabilities of larger serrations. In contrast, a significant small-scale tail can be seen for the initial sample in the inset of Fig. 9(a), conforming to the negative skewness of the distribution.

To complete the consideration of the instabilities manifested on the deformation curve, it should be recalled that besides large stress drops, some papers attracted attention to much smaller stress fluctuations, corresponding to a separate scale range, which may occur during jerky flow at low enough $\dot{\varepsilon}_a$ [36,37,43]. These fluctuations were shown to obey a physically distinct statistical law. Specifically, the probability to find a stress drop with an amplitude δσ was found to scale as a power of δσ [40]: $P(δσ) \sim δσ^{-α}$. Such power-law statistics, which describe scale-free behavior, are known to be a ubiquitous feature of small-scale deformation processes, e.g., captured by virtue of AE or local extensometry measurements both during jerky flow in alloys [38,43] and smooth flow in pure materials [41,62]. It was suggested that the collective dislocation motions relevant to the corresponding scale ranges are governed by dynamical laws different from those controlling the macroscopic jerky flow. Namely, the scale invariance is commonly considered as a signature of an avalanche nature of the dislocation dynamics. Generally, the interpretation involves the phenomenon of self-organized criticality (SOC) suggested as a paradigm of earthquakes [23]. An alternative concept suggested a mechanism of emergence of a power-law scaling in terms of turbulent flows [63].

Small serrations with amplitudes below 0.1 MPa were indeed found for all samples in the presented experiments. As mentioned above, these fluctuations were not explicitly related to specific strain localizations. It may be suggested that they represent a summary force response to small strain-rate bursts taking place in different locations. Numerous small serrations occurred in the region of the elastoplastic transition (cf. [64]), but their number was drastically reduced afterword, thus making difficult a quantitative comparison of the statistics for different samples. Nevertheless, the analysis in the amplitude range from 0.1 MPa down to the noise level, which covers about 1.5 orders of magnitude, clearly revealed power-law behavior in all cases. The exponent α varied approximately between 1.5 and 2, which agrees with the typical values reported in the literature, as synthesized in [50].

### 3.4. Effect on the small-scale complexity: DIC and AE responses

A further insight into phenomena at smaller scales was gained using DIC and AE methods. Fig. 10 illustrates the main qualitative result obtained due to DIC measurements. The figure presents magnified and deliberately overcontrasted portions of spatiotemporal maps which make appear the above-mentioned heterogeneity of plastic flow outside the PLC bands. Hot-color spots represent local strain-rate bursts ("events") reflecting intermittent deformation processes with a lower intensity than the PLC bands. The bursts are also accompanied with unloading in the neighbourhood due to the springback effect, which appears as blue spots. The spots have a height varying from one grid step (about 0.1 mm) to typically less than ten steps. Therewith, the intermittent processes last very shortly, so that the apparent spot duration is determined by the sampling time of 0.1 s. Control tests with a higher acquisition rate proved that the burst durations were less than 0.02 s. It should also be specified for clarity that the PLC bands modify these patterns in two ways. First, they are often accompanied with visible unloading sequences in multiple sites throughout the entire gauge length. Besides, the PLC bands are regularly preceded and succeeded by an increased local activity, often in the form of tails composed of local spots, which correspond to the propagation of plastic flow. Such patterns are well seen in Fig. 10(b) where PLC bands appear fully within the gauge length.

Fig. 10(a) demonstrates that the local strain-rate field corresponding to the macroscopically smooth plastic flow in the initial material is presented by a scatter of separate events. Substantially, this pattern reveals that the local events tend to arrange along inclined lines testifying to the propagation of plastic activity. Such patterns may be an intrinsic property of plastic flow because similar observations were reported not only for a similar Al-Mg alloy [38], but also for other alloys and even pure materials, both single and polycrystals [58,62]. The propagation velocity is rather steady in the present example and slightly varies about 8 mm/s, but noticeable variations by a factor up to about 2 were also observed. Plots (b) and (c) represent the effect of SMAT on such maps. It occurred that in contrast to the macroscopic characteristics studied above, the changes in the fine-scale patterns do not mark out the 2L case but may be rather grouped according to the symmetry of the treatment. When two surfaces were processed, i.e., in the 2N and 2L cases, the overall density of spots decreased, they became less ordered and the tendency to their propagation less pronounced, as illustrated in Fig. 10(b) for a 2L sample. At the same time, the pattern remained qualitatively similar to that observed in the initial state that also corresponds to symmetrical conditions for the opposite faces of the specimen. The effect of SMAT drastically changed when only one surface was treated, i.e., in the 1N and 1L cases, as displayed in Fig. 10(c). It can be seen that the local activity of plastic flow is strongly delocalized in the sense that it synchronously covers large portions of the gauge length at the same time instants, as reflected in the vertical arrangement of the spots. Some spots are aligned along steep lines corresponding to the propagation of plastic activity with a high velocity, as depicted by dashed lines. It may thus be suggested that the vertical segments correspond to the limit case of a very high velocity of propagation of an "excitation" of plastic activity along the specimen.

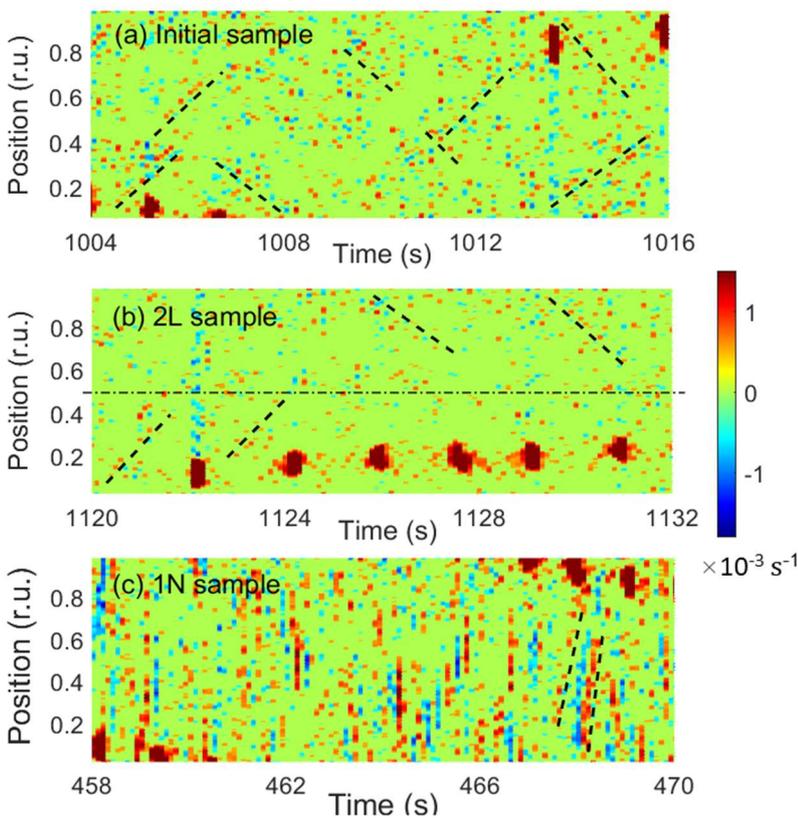

*Figure 10.* Magnified portions of spatiotemporal maps for (a) an initial sample, (b) a 2L sample, and (c) 1N sample. For convenience, dashed lines point to some of the propagation sequences. The horizontal dash-dotted line indicates the trace of the $\dot{\varepsilon}_{loc}(t)$ signal used for statistical analysis.

Basically, both observed influences are conforming to the delocalization effect of SMAT remarked above for the macroscopic scale of the PLC bands. In the case of two processed surfaces this tendency is related to a suppression of the local strain-rate bursts, i.e., a more uniform plastic flow. In contrast, the treatment of one surface intensifies the small-scale deformation processes. However, the delocalization occurs in this case too, as a fast synchronization of local bursts along the gauge length, a kind of a switching wave [65].

These qualitative observations are corroborated by the statistical analysis of the amplitudes of fluctuations of the local strain rate at one point selected in the middle of the gauge length. Such a signal corresponds to a horizontal cross-section of the spatiotemporal map as depicted by a dash-dotted line in Fig. 10(b). The analysis met a difficulty caused by a constraint of the variation of the fluctuation amplitudes to a narrow scale range. On the one hand, the lower signal level was significantly limited by the threshold selected to fully cut off the noise, including the outliers. On the other hand, an upper threshold was additionally used to cut the bursts associated with the PLC bands and also their precursors and aftershocks. The remaining range of amplitudes was slightly narrower than one order of magnitude. Consequently, although the resulting distributions displayed descending dependences, it was not possible to distinguish reliably a power law from an exponential decay. Nevertheless, the statistical analysis of the entire signal below the upper threshold, i.e., with the noise included, revealed a quantitative expression of the above-described qualitative comparison of different patterns. Fig. 11(a) presents such histograms of distribution of the fluctuation amplitudes for an initial sample and two specimens corresponding to a different symmetry of SMAT. All histograms have a peaked shape with a similar left part which is likely to be determined by the noise. However, the right branch, which contains both a part of noise and the fluctuations caused by the deformation processes, is considerably changed after SMAT. Namely, this branch is compressed after SMAT of two surfaces, in agreement with the depletion of the ensemble of bursts in the spatiotemporal map in Fig. 10(b), while the SMAT of one surface leads to an increased intensity of fluctuations [cf. Fig. 10(c)]. Fig. 11(b) depicts the respective changes in the mean amplitude of the fluctuations, which confirm the above statements.

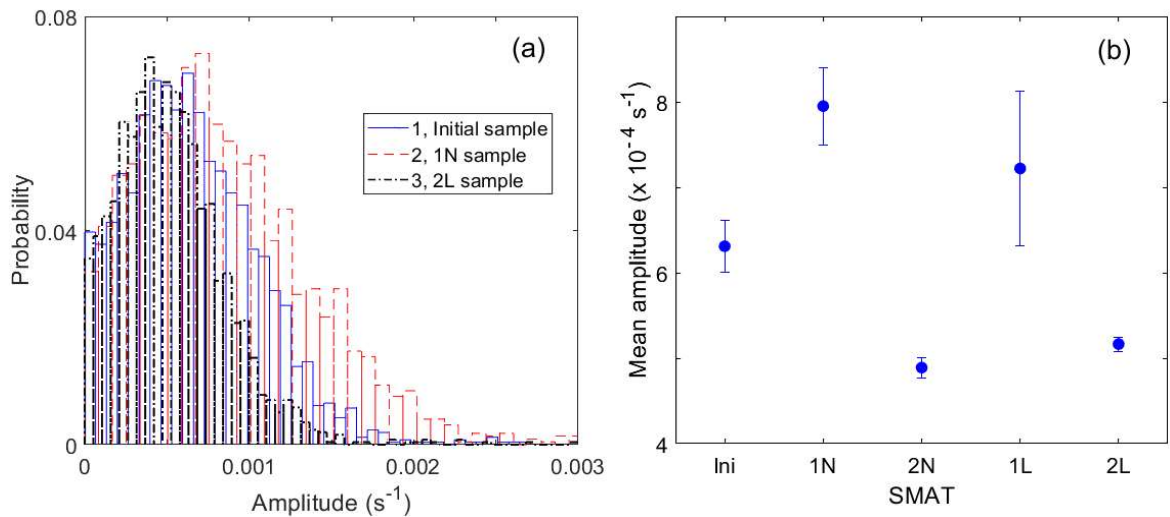

*Figure 11. SMAT effect on the fluctuations of the local strain rate, $\dot{\varepsilon}_{loc}(t)$. Each signal was extracted for a point selected in the middle of the gauge area, as illustrated in Fig. 10(b). (a) Examples of statistical distributions of the fluctuation amplitudes: 1 – initial sample; 2 – 1N sample; 3 – 2L sample. (b) Average amplitude of fluctuations for various treatment modes.*

These results were completed by an AE study which provides a complementary view of the dynamics of deformation processes at the least accessible scale. Indeed, on the one hand the AE technique assures a time resolution at least four orders of magnitude better than the DIC and, theoretically, is able to feel the breakthrough of individual pile-ups [66]. On the other hand, the records realized using a single AE sensor represent a nonlocal response of the material, which gathers sound waves from the entire specimen bulk. Figs. 12(a-c) illustrate the results of AE measurements for a 1L sample by putting into correspondence the stress-time curve, the logarithmic amplitudes of the AE events, $U$, and the cumulated amplitude, $A_{cum}$, obtained as a sum of linear amplitudes of the events recorded up to the given time instant. The depicted AE behavior is similar to the previous observations on Al-Mg alloys [43,50] and common to various metallic materials, as illustrated in Fig. 12(b) using a $U(t)$ dependence for the 1L case. Namely, AE is intense during the elastoplastic transition due to a fast multiplication and a large free path of dislocations, but gradually decreases with the material work-hardening [67]. Therewith, the relative similitude between σ(t) and $A_{cum}(t)$ proves that a considerable part of plastic flow gives rise to a measurable AE. With this analogy in view, the statistical analysis was performed in the spirit of the previous works [43,50]. Accordingly, the squared amplitude of events, $A^2$, was considered as a measure of the energy dissipated in the underlying deformation process [68]. Fig. 12(d) shows that the AE obeyed power-law amplitude distributions both in the initial state and after SMAT, which agrees with the general belief that the dislocation dynamics is intrinsically avalanche-like at the scale pertaining to the AE [41,43].

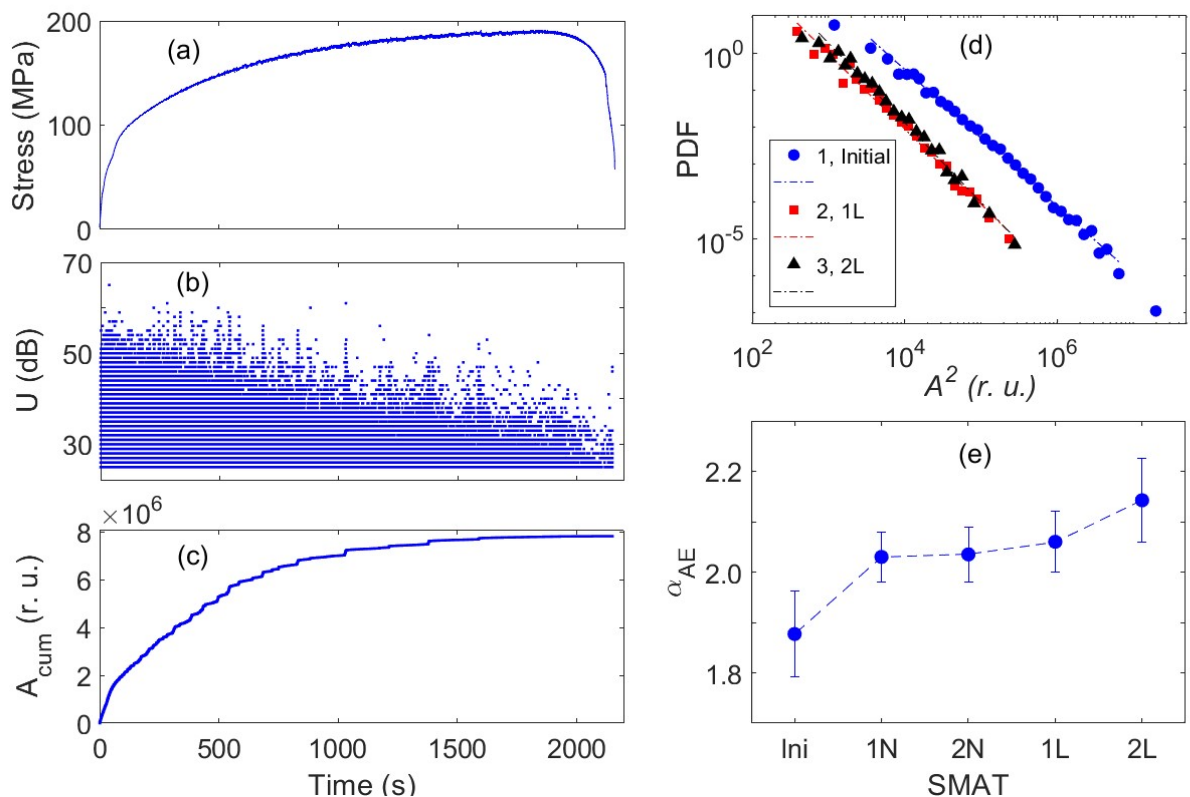

*Figure 12.* SMAT effect on the AE. The example of AE measurements is illustrated for a 1L sample: (a) Deformation curve, (b) logarithmic amplitude, U, of the accompanying AE events, and (c) cumulated sum, $A_{cum}$, of the corresponding linear amplitudes of AE events (relative units). (d) Examples of the probability density function, PDF, for squared amplitudes, $A^2$, of AE events: 1 – Initial sample, 2 – 1L sample, and 3 – 2L sample. (e) Power-law exponent, $α_{AE}$, determined as the value of the slope of the dependences such as depicted in Fig. 12(d).

The comparison between different samples revealed that the influence of SMAT on the plastic flow is visible at the AE scale. First, the AE intensity was maximum in the initial materials and demonstrated a tendency to an overall decrease after SMAT. Furthermore, fig. 12(d) testifies that this tendency did not only concern the average AE intensity but also the events repartition relative to their amplitudes. Indeed, the probability distribution function obtained for an initial sample corresponds to a lower exponent $\alpha_{AE}$, which indicates a relatively higher probability of larger events than for SMAT samples. Although $\alpha_{AE}$ evolved with deformation and showed a tendency to increase with strain before the onset of necking (cf. [43]), this inequality persisted at all stages of work hardening. As an example, Fig. 12(e) presents $\alpha_{AE}$ values obtained after different treatments for a strain interval where the AE was approximately stationary for every sample. The data display a tendency to a growth in $\alpha_{AE}$ after SMAT, which reflects a higher probability of smaller events. Interestingly, this tendency is opposite to that discussed for the macroscopic stress serrations, as followed from the analysis of the peaked shapes of $\Delta\sigma$ distributions. However, a direct comparison of behaviors at two distinct scales would be misleading because the AE statistics includes all deformation processes giving rise to an acoustic response, during both short stress serrations and longer reloading periods.

## 4. Discussion

The results presented in Sections 3.2 and 3.3 clarify that surface modifications by SMAT affected various mechanical properties of the studied Al-Mg alloy. This includes the elastoplastic transition, the strain hardening, as well the PLC and Lüders instabilities which represent macroscopic deviations from a uniform plastic flow in the form of transitory and propagating deformation bands. Since the macroscopic instabilities attest to a striking importance of the dislocations self-organization, which involves deformation processes on multiple scales, mechanical responses pertaining to finer deformation processes were investigated using DIC and AE methods (Sec. 3.4). It appeared that both the local strain-rate fields and the AE were also affected by the surface conditions.

The results obtained bear witness that the changes induced by SMAT can be manifested at different scales and depend on the specimen gauge surfaces treated. This complexity implies consideration of diverse factors that may have different consequences on the plastic flow. The variety of the observations can be qualitatively addressed by considering three competing effects. On the one hand, (i) the introduction of a high dislocation density in the surface layers (see Figs. 1 and 2) brings about new sources of mobile dislocations [69]. This aspect may explain, e.g., an early onset of microplasticity in SMAT samples, as evidenced by the decrease in the proportionality limit in comparison with the initial material. In its turn, relaxation of the local stresses via microplasticity may contribute to the suppression of the Lüders plateau (Figs. 3 and 5) and the increases in the critical stresses for the onset of both small irregular stress serrations and the PLC effect (Fig. 6) [37,70]. On the other hand, as the forest dislocations are important obstacles to the dislocation motion in *fcc* metals, (ii) the increase in the dislocation density will reduce their mobility in the surface layer. This factor may be an alternative mechanism preventing the occurrence of small serrations by hindering the progress of weak deformation bands [37]. The surface hardening may also be called to explain the initial increase in the deforming stress (Fig. 3). However, the competition of these two opposite effects, one softening and the other hardening of the surface layers, do not suffice to answer why SMAT can lead to qualitatively different effects depending on the treatment geometry (see, e.g., Figs. 5, 9, 10 and 11). It should be included that the surface hardening by SMAT produces an initial difference in plasticity between the hard surface layers and the softer bulk, which implies (iii) a back-stress strengthening effect [71]. More specifically, the plastic deformation will start earlier in the bulk, but its progress will be constrained by the surface layers. The resulting strain incompatibility will lead to GND generation in the intermediate layers, thus building up a complex stress state and involving more slip systems into plastic flow. Such a process is usually considered as a mechanism allowing for a high strength combined with a good ductility of materials with gradient microstructures [35,56]. It is also associated with the up-turn in the deformation curves, typical of such materials (cf. Fig. 3). At the same time, spatiotemporal maps presented in Figs. 5(b-d) revealed that depending on the choice of the surface(s) processed, the increase in the work-hardening rate, which causes this up-turn, may correspond to qualitatively distinct patterns of the local strain heterogeneity. It seems instinctive in this context that the constraints induced by the strengthening of two large surfaces resulted in a highly uniform spatial pattern before the onset of the PLC effect (Fig. 5) and a virtual absence of the initial type *A* behavior upon the onset of the instability, as compared with the "medium" treatments that leave intact at least one large surface. It might be recalled in this context that type *A* behavior is usually observed in annealed alloys in similar deformation conditions and transforms into type *B* after some strain hardening [14]. Overall, the fact that 2L SMAT is the only case when both large surfaces are impacted may explain the particular position of this treatment, as manifested in Figs. 3, 5, 6 and 9.

Several more observations deserve special attention. As far as the qualitative features of the PLC effect are concerned, the dominant type *B* character of stress serrations and the hopping propagation of deformation bands persisted against all kinds of surface treatment (Fig. 3 and 7). This result agrees with various literature data which proved robustness of the type *B* regime in a large interval of strain rates around $10^{-4}$ s$^{-1}$ for Al alloys with various initial microstructures [7,72,73]. Also, the same mechanism of the band formation, namely, a virtually simultaneous nucleation of a band across the specimen [21], became dominant in all materials after some deformation. This robustness allowed for a quantitative comparison of the PLC effect in a strain interval corresponding to stabilized macroscopic serrations. One of the major results of this analysis is that despite the quickly established similar patterns of behavior, the serration amplitude and frequency as well as the PLC band width remained sensitive to the surface pre-deformation at all strains (Fig. 9). The significant changes in these characteristics after medium SMAT are likely to be caused by a combination of the factors (i) and (ii). Indeed, introduction of a great number of dislocation nucleation sites may be responsible for a more frequent triggering of instability and formation of wider deformation bands. The possible role played by this preset heterogeneity is corroborated by the analysis of the higher-order statistical moments of the amplitude distributions, which signaled the enhanced number of large outliers after SMAT. On the other hand, the reduced dislocation mobility would limit the strain rates within the bands and, consequently, the serration amplitudes. Surprisingly, these two effects roughly compensated each other, which resulted in a similar contribution of plastic instability to the total strain in all cases. Moreover, calculation of the cumulated amplitude of stress drops over the entire deformation curves until the onset of necking conformed this conclusion. Such a constancy presents an interesting question for theoretical modeling of the PLC effect.

It is also noteworthy that the influence of SMAT on the PLC band width evolution corroborates the conjecture that consideration of a spatial coupling via internal stresses within a heterogeneously strained material is indispensable for understanding the collective deformation processes and predicting behavior of the PLC effect in real alloys (see, e.g., [9-13]). In contrast to the PLC band velocity that is known to diminish with strain [33,47,60], there is no consensus in the literature on the strain dependence of the band width. Both an increasing and decreasing trends have been reported by various authors [60,74]. In the present work, both these trends were observed depending on the surface state, as clarified in Fig. 9(b). The following intuitive considerations can be suggested. The greater initial band width in the untreated samples is likely to be due to their high microstructure uniformity, so that a local instability event may trigger instability in multiple neighboring sites that find themselves in a similar stress state. The further band narrowing may be caused by the progressive reduction of the dislocation mobility and also by building up the heterogeneity caused by the previous deformation bands. SMAT creates an initially heterogeneous surface state, which would result in narrower deformation bands. However, this heterogeneity itself might promote wider bands as deformation progresses and the overall level of internal stresses is increased.

Perhaps, the most striking argument in favor of the above hypotheses on the mechanisms associated with the surface pre-deformation was provided by the behavior of the fine-scale heterogeneity of the local strain-rate field detected by DIC outside the PLC bands (Figs. 10 and 11). The major observation concerns the ubiquitous character of correlated behavior of the local strain-rate bursts, manifested as waves of plastic activity [75]. Such a correlation proves the spatial coupling of deformation processes at the relevant scale (cf. [62]). Therewith, the extremely fast propagation of plastic activity along the samples with one surface treated by SMAT confirms the high density of dislocation sources after the treatment, while the destruction of the spatial correlation after treatment of two opposite surfaces agrees with the suggestion that the operation of the surface dislocation sources may be impeded by the back stresses.

The spatial coupling is also manifested in the occurrence of precursors and aftershocks – the tails of local strain-rate bursts preceding or following the PLC bands (Fig. 10). This feature adds an interesting piece of information to the question on the evolution of the individual PLC bands. Indeed, while the presence of "precursors" triggering a macroscopic instability seems natural and has been observed, e.g., using the AE technique [43], the possibility of "aftershocks" at the bottom of the stress drop has not been discussed so far.

In view of the complex nature of the spatial coupling, it is important to underline that the application of the methods pertaining to different scales of deformation processes revealed that the apparent behaviors may depend on the scale of observation. Indeed, the behaviors change from avalanche-like processes, as reflected in the scale-invariant statistics of the AE and of small stress serrations occurring all over the deformation curves, to propagation of local strain bursts revealed by the DIC, and finally to macroscopic stress serrations caused by the PLC bands. A hint to these changes may be found in the analysis of the AE accompanying the PLC serrations (e.g., [76]). This investigation led to a conjecture that a macroscopic stress drop results from a dense chaining of multiple dislocation avalanches, while the power-law statistics of the AE agrees with the concept of a ubiquitously avalanche-like nature of deformation processes at a microscopic scale. It may be suggested that the strain-rate bursts accessible to DIC observations are also governed by such chained processes but involve much less avalanches breaking in local regions when the strain/stress state of the material is far from the threshold for the macroscopic instability.

In this connection, two additional remarks are worthwhile from the quantitative viewpoint. First, the analysis of the AE statistics allowed to detect an increase in the power-law exponent after SMAT [Figs. 12(d) and 12(e)], which attests to a higher probability of weaker events. This trend corroborates the conjecture that one of the factors that must be considered to interpret the SMAT effect on the plastic flow is the decrease in the dislocation mobility. Second, the fluctuations of the local strain rate, which correspond to an intermediate scale between the stress drops and the AE, do not obey a power law (Fig. 12). This result agrees with the conclusion made for a high-entropy Al alloy using a similar DIC method [48]. However, it contradicts the results obtained by a one-dimensional (1D) method of measurement of the local strain rate in an Al-Mg alloy [38] and pure Cu single crystals [61]. Since the 1D technique provided a considerably better frequency and strain resolution at the expense of a small number of surveyed local sites, the interpretation of the statistics obtained by the DIC method remains an open question.

## 5. Conclusions

In summary, analysis of plastic deformation at three distinct scales of deformation processes allowed to obtain significant information on the influence of the surface pre-deformation by SMAT on mechanical behavior and PLC effect in a model Al-Mg alloy. One of the findings at the scale of the deformation curve, which may be useful from the practical viewpoint, is that SMAT can increase the macroscopic yield strength without involving a decrease in the ductility of the material. The ductility preservation is likely to be associated with a progressive leveling of strengthening between surface and bulk of the material. Nevertheless, the influence of SMAT on the PLC effect persists during deformation, as proven by the analysis of stress serrations and deformation band characteristics. This analysis led to an important fundamental hypothesis of delocalization of the PLC bands in SMAT specimens in comparison with the untreated material. Moreover, DIC measurements revealed that this conjecture also concerns a finer scale of local strain-bursts outside the PLC bands.

The results obtained by DIC also confirmed the recently suggested hypothesis of a wave-intermittence duality of collective deformation processes at the respective scale [62]. On the one hand, the local strain-rate field is not uniform and displays local bursts above the level prescribed by the applied strain rate. Such intermittence is conforming to the concept of intrinsic avalanche behavior of dislocations, because an essential part of the AE accompanying plastic deformation of various materials obeys power-law statistics. On the other hand, the sequences of such strain-rate bursts, as observed in spatiotemporal maps, may be arranged in wavy structures indicating the propagation of these events along the specimen length. Importantly, in contrast to the initial belief that such patterns have a universal character and may be similar in different materials, the data obtained in the present paper prove that they are sensitive to the surface state and, therefore, may be dependent on the microstructure.

To interpret the entirety of the results, a competition of several possible consequences of the surface modification was considered, including the generation of dislocations sources as well as creation of obstacles to the motion of dislocations, but also an asymmetry between the hardened surface and the soft bulk, which results in back-stress hardening. The competition of several factors played by the surface in the plastic flow related to different scales creates a sophisticated problem. Indeed, the subtle balance between different trends can be shifted by a variation in the experimental conditions, in particular, in the composition or microstructure of the alloy. This complexity presents a challenge for modelling plastic flow of crystals at different scales. In particular, the data obtained bear evidence that the realistic models of the PLC instability in alloys will benefit from the consideration of deformation processes on mesoscopic scales.

**Acknowledgements**

M.L. and T.L gratefully acknowledge the support from the French State through the program "Investment in the future" operated by the National Research Agency, in the framework of the LabEx DAMAS [ANR-11-LABX-0 0 08–01], and the research program RESEM managed by IRT M2P (Institut de Recherche Technologique en Matériaux, Métallurgie et Procédés). The authors acknowledge the experimental facilities MécaRhéo, MicroMat and Procédés from the LEM3 (Université de Lorraine - CNRS UMR 7239).

**References**

[1] A. Portevin, F. Le Chatelier, Sur un phénomène observé lors de l'essai de traction d'alliages en cours de transformation, C.R. Acad. Sci. Paris. 176 (1923) 507–510.

[2] K. Lu, J. Lu, Nanostructured surface layer on metallic materials induced by surface mechanical attrition treatment, Mater. Sci. Eng. A375-377 (2004) 38–5.

[3] T.H. Fang, W.L. Li, N.R. Rao, K.Lu, F. Savart, Revealing Extraordinary Intrinsic Tensile Plasticity in Gradient Nano-Grained Copper, Science 331 (2011) 1587–1590.

[4] T.O. Olugbade, J. Lu, Literature review on the mechanical properties of materials after surface mechanical attrition treatment (SMAT), Nano Mater. Sci. 2 (2020) 3–31.

[5] Yu.A. Filatov, V.I. Yelagin, V.V. Zakharov, New Al-Mg-Sc alloys, Mater. Sci. Eng. A280 (2000) 97–101.

[6] P. Bazarnik, M. Lewandowska, K.J. Kurzydlowskp, Mechanical behaviour of ultrafine grained Al-Mg alloys obtained by different processing routes, Arch. Metall. Mater. 57 (2012) 869–876.

[7] A. Mogucheva, D. Yuzbekova, R. Kaibyshev, T. Lebedkina, M. Lebyodkin, Effect of grain refinement on jerky flow in an Al-Mg-Sc alloy, Metall. Mater. Transact. A47 (2016) 2093–2106.

[8] S. Chen, J. Wang, L. Xia, Y. Wu, Deformation behavior of bulk metallic glasses and high entropy alloys under complex stress fields: a review, Entropy 21 (2019) 54.

[9] M.A. Lebyodkin, T.A. Lebedkina, J. Brechtl, P.K. Liaw, Serrated Flow in Alloy Systems, in: J. Brechtl, P.K. Liaw (Eds.), High-Entropy Materials: Theory, Experiments, and Applications, Springer International Publishing, Cham, 2021, pp. 523–644.

[10]  P.G. McCormick, Theory of flow localization due to dynamic strain ageing, Acta metall. 36 (1988) 3061–3067.

[11]  Y. Estrin, L.P. Kubin, Collective dislocation behavior in dilute alloys and the Portevin-Le Chatelier effect, J. Mech Behav. Mater. 2 (1990) 255–292.

[12]  M. Mazière, J. Besson, S. Forest, B. Tanguy, H. Chalons, F. Vogel, Numerical aspects in the finite element simulation of the Portevin–Le Chatelier effect, Comput. Methods Appl. Mech. Engrg. 199 (2010) 734–754.

[13]  S. Tamimi, A. Andrade-Campos, J. Pinho-da-Cruz, Modelling the Portevin-Le Chatelier effects in aluminium alloys: a review, J. Mech. Behav. Mater. 24 (2015) 67–78.

[14]  P. Rodriguez, Serrated plastic flow, Bull. Mater. Sci. 6 (1984) 653–663.

[15]  D.A. Zhemchuzhnikova, M. Lebyodkin, T. Lebedkina, R.O. Kaibyshev, Unusual behavior of the Portevin-Le Chatelier effect in an AlMg alloy containing precipitates, Mater. Sci. Eng. A 639 (2015) 37–41.

[16]  M.V. Markushev, M.Yu. Murashkin, Structure and mechanical properties of commercial Al–Mg 1560 alloy after equal-channel angular extrusion and annealing, Mater. Sci. Eng. A367 (2004) 234–242.

[17]  G. Saad, S.A. Fayek, A. Fawzy, H.N. Soliman, E. Nassr, Serrated flow and work hardening characteristics of Al-5356 alloy, J. Alloy. Compd. 502 (2010) 139–146.

[18]  K. Chihab, Y. Estrin, L.P. Kubin, J. Vergnol, The kinetics of the Portevin-Le Chatelier bands in an Al-5 at%Mg alloy, Scr. Metall. 21 (1987) 203–208.

[19]  H. Ait-Amokhtar, P. Vacher, S. Boudrahem, Kinematics fields and spatial activity of Portevin-Le Chatelier bands using the digital image correlation method, Acta Mater. 54 (2006) 4365–4371.

[20]  S. Bakir, Instabilité plastique propagative liée au phénomène Portevin-Le Chatelier dans les alliages aluminium-magnésium, Doctoral thesis, Université de Metz, France (1995).

[21]  M.A. Lebyodkin, D.A. Zhemchuzhnikova, T. Lebedkina, E.C. Aifantis, Kinematics of formation and cessation of type B deformation bands during the Portevin-Le Chatelier effect in an AlMg alloy, Results Phys. 12 (2019) 867–869.

[22]  L.P. Kubin, C. Fressengeas, G. Ananthakrishna, Collective behaviour of dislocations, Dislocations in solids 11 (2002) 101–192.

[23]  P. Bac, C. Tang, K. Wiesenfeld, Self-organized criticality, Phys. Rev. A38 (1988) 364–374.

[24]  H.D.I. Abarbanel, R. Brown, J.J. Sidorowich, L.S. Tsimring, The analysis of observed chaotic data in physical systems, Rev. Mod. Phys. 65 (1993) 1331–1392.

[25]  S.H. Strogatz, From Kuramoto to Crawford: exploring the onset of synchronization in populations of coupled oscillators, Physica D 143 (2000) 1–20.

[26]  Y. Estrin, A. Vinogradov, Extreme grain refinement by severe plastic deformation: A wealth of challenging science, Acta Mater. 61 (2013) 782–817.

[27]  R.Z. Valiev, B. Straumal, T.G. Langdon, Using severe plastic deformation to produce nanostructured materials with superior properties, Ann. Rev. Mater. Res. 52 (2022) 357–82.

[28]  M. Jamalian, D.P. Field, Effects of shot peening parameters on gradient microstructure and mechanical properties of TRC AZ31, Mater. Charact. 148 (2019) 9–16.

[29] J. Ding, Q. Li, J. Li, S. Xue, Z. Fan, H. Wang, X. Zhang, Mechanical behavior of structurally gradient nickel alloy, Acta Mater. 149 (2018) 57–67.

[30] R. Kalsar, S. Suwas, A novel way to enhance the strength of twinning induced plasticity (TWIP) steels, Scr. Mater. 154 (2018) 207–211.

[31] L. Casarotto, H. Dierke, R. Tutsch, H. Neuhäuser, On nucleation and propagation of PLC bands in an Al–3Mg alloy, Mater. Sci. Eng. A527 (2009) 132–140.

[32] A.A. Shibkov, M.A. Zheltov, M.F. Gasanov, A.E. Zolotov, A.A. Denisov, and M.A. Lebyodkin, Dynamics of deformation band formation investigated by high-speed techniques during creep in an AlMg alloy, Mater. Sci. Eng. A 772 (2020) 138777.

[33] D. Yuzbekova, A. Mogucheva, Yu. Borisova, R. Kaibyshev, On the mechanisms of nucleation and subsequent development of the PLC bands in an AlMg alloy, J. Alloy. Compd. 868 (2021) 159135.

[34] X. Meng, B. Liu, L. Luo, Y. Ding, X.-X. Rao, B. Hu, Y. Liu, J. Lu, The Portevin-Le Châtelier effect of gradient nanostructured 5182 aluminum alloy by surface mechanical attrition treatment, J. Mater. Sci. Technol. 34 (2018) 2307–2315.

[35] H. Hassanpour, R. Jamaati, S.J. Hosseinipour, Effect of gradient microstructure on the mechanical properties of aluminum alloy, Mater. Charact. 174 (2021) 111023.

[36] D. Thevenet, M. Mliha-Touati, A. Zeghloul, The effect of precipitation on the Portevin-Le Chatelier effect in an Al-Zn-Mg-Cu alloy, Mater. Sci. Eng. A266 (1999) 175–182.

[37] N.P. Kobelev, M.A. Lebyodkin, T.A. Lebedkina, Role of Self-Organization of Dislocations in the Onset and Kinetics of Macroscopic Plastic Instability, Metall. Mater. Trans. A-Phys. Metall. Mater. Sci. 48A (2017) 965–974.

[38] M. Lebyodkin, Y. Bougherira, T. Lebedkina, D. Entemeyer, Scaling in the Local Strain-Rate Field during Jerky Flow in an Al-3%Mg Alloy, Metals 10 (2020) 134.

[39] F.F. Csikor, Ch. Motz, D. Weygand, M. Zaiser, S. Zapperi, Dislocation Avalanches, Strain Bursts, and the Problem of Plastic Forming at the Micrometer Scale, Science 318 (2007) 251–254.

[40] D.M. Dimiduk, C. Woodward, R. LeSar, M.D. Uchic, Scale-Free Intermittent Flow in Crystal Plasticity, Science 312 (2006) 1188–1190.

[41] J. Weiss, W.B. Rhouma, T. Richeton, S. Deschanel, F. Louchet, L. Truskinovsky, From mild to wild fluctuations in crystal plasticity, Phys. Rev. Lett. 114 (2015) 105504.

[42] A.Yu. Vinogradov, D.L. Merson, The Nature of Acoustic Emission during Deformation Processes in Metallic Materials, Low Temp. Phys. 44 (2018) 930–937.

[43] M.A. Lebyodkin, N.P. Kobelev, Y. Bougherira, D. Entemeyer, C. Fressengeas, V.S. Gornakov, T.A. Lebedkina, I.V. Shashkov, On the similarity of plastic flow processes during smooth and jerky flow: statistical analysis, Acta Mater. 60 (2012) 3729–3740.

[44] W. Pantleon, Resolving the geometrically necessary dislocation content by conventional electron backscattering diffraction, Scr. Mater. 58 (2008) 994–997.

[45] B. Beausir, J.J. Fundenberger, Analysis Tools for Electron and X-ray diffraction, ATEX-Software, www.Atex-Software, Eu, Université de Lorraine-Metz, France, 2017.

[46] M.A. Sutton, F. Hild, Recent Advances and Perspectives in Digital Image Correlation, Exp. Mech. 55 (2015) 1–8.

[47] D. Yuzbekova, A. Mogucheva, D. Zhemchuzhnikova, T. Lebedkina, M. Lebyodkin, R. Kaibyshev, Effect of microstructure on continuous propagation of the Portevin–Le Chatelier deformation bands, Int. J. Plast. 96 (2017) 210–226.

[48] J. Brechtl, R. Feng, P.K. Liaw, H. Jaber, T. Lebedkina, M. Lebyodkin, Mesoscopic-Scale Complexity in Macroscopically-Uniform Plastic Flow of an $Al_{0.3}CoCrFeNi$ High-Entropy Alloy, Acta Mater. 242 (2023) 118445.

[49] http://www.correlatedsolutions.com/vic-2d

[50] T.A. Lebedkina, D.A. Zhemchuzhnikova, M.A. Lebyodkin, Correlation versus randomization of jerky flow in an AlMgScZr alloy using acoustic emission, Phys. Rev. E 97 (2018) 013001.

[51] N.N. Hsu, F.R. Breckenridge, Characterization of acoustic emission sensors, Mater. Eval. 39 (1981) 60–68.

[52] L. Adam, E. Bonhomme, C. Chirol, A. Proust, Benefits of acoustic emission for the testing of aerospace composite assemblies, 30th European Conference on Acoustic Emission Testing & 7th International Conference on Acoustic Emission University of Granada, 12-15 September 2012, www.ndt.net/EWGAE-ICAE2012/.

[53] M.A. Lebyodkin, T.A. Lebedkina, F. Chmelík, T. Lamark, Y. Estrin, C. Fressengeas, and J. Weiss, Intrinsic structure of acoustic emission events during jerky flow in an Al alloy, Phys. Rev. B 79 (2009) 174114.

[54] E. Pomponi, A. Vinogradov, A. Danyuk, Wavelet based approach to signal activity detection and phase picking: application to acoustic emission, Signal Processing 115 (2015) 110–119.

[55] E.O. Hall, Yield point phenomena in metals and alloys, Springer, NY, USA, 2012.

[56] X. Wua, P. Jianga, L. Chena, F. Yuana, Y.T. Zhu, Extraordinary strain hardening by gradient structure, Proc. Natl. Acad. Sci. U. S. A. 111 (2014) 7197–7201.

[57] S.-Y. Lee, S. Chettri, R. Sarmah, C. Takushima, J.-I. Hamada, N. Nakada, Serrated flow accompanied with dynamic type transition of the Portevin–Le Chatelier effect in austenitic stainless steel, J. Mater. Sci. Techn. 133 (2023) 154–164.

[58] A. Roth T.A. Lebedkina, M.A. Lebyodkin, On the critical strain for the onset of plastic instability in an austenitic FeMnC steel, Mater. Sci. Eng. A 539 (2012) 280–284.

[59] L. Kubin, Y. Estrin. Dynamic strain ageing and the mechanical response of alloys, J. Phys. III 1 (1991) 929–943.

[60] R. Shabadi, S. Kumar, H. J. Roven, E.S. Dwarakadasa, Effect of specimen condition, orientation and alloy composition on PLC band parameters, Mater. Sci. Eng. A 382 (2004) 203–208.

[61] R.A. Groenveld, G. Meeden, Measuring Skewness and Kurtosis, The Statistician 33 (1984) 391–399.

[62] C. Fressengeas, A.J. Beaudoin, D. Entemeyer, T. Lebedkina, M. Lebyodkin, V. Taupin, Dislocation transport and intermittency in the plasticity of crystalline solids, Phys. Rev. B 79 (2009) 014108.

[63] G. Ananthakrishna, M.S. Bharathi, Dynamical approach to the spatiotemporal aspects of the Portevin–Le Chatelier effect: Chaos, turbulence, and band propagation, Phys. Rev. E 70 (2004) 026111.

[64] R.N. Mudrock, M.A. Lebyodkin, P. Kurath, A.J. Beaudoin, T.A. Lebedkina, Strain-rate fluctuation during macroscopically uniform deformation of a solution-strengthened alloy. Scr. Mater. 65, 1093–1096 (2011).

[65] J. Lottes, G. Biondini, S. Trillo Excitation of switching waves in normally dispersive Kerr cavities, Optics Letters 46 (2021) 2481–2484.

[66] K. Mathis, F. Chmelik, Exploring plastic deformation of metallic materials by the acoustic emission technique, in Acoustic Emission, ed. by W. Sikorsky, InTech, Croatia, 2012, pp. 23–48.

[67] C. R. Heiple, S.H. Carpenter, Acoustic emission produced by deformation of metals and alloys – a review, Part I, J. Acoustic Emission 6 (1987) 177–204.

[68] J. Weiss, J.-R. Grasso, Acoustic emission in single crystals of ice, J. Phys. Chem. B 101 (1997) 6113.

[69] H. Zhou, C. Huang, X.o Sha, L. Xiao, X. Ma, H.W. Höppel, M. Göken, X. Wu, K. Ameyama, X. Han, Y. Zhu, In-situ observation of dislocation dynamics near heterostructured interfaces, Mater. Res. Lett. 7 (2019) 376–382.

[70] P.G. McCormick, Strain rate sensitivity prior to the onset of serrated yielding in a pressurized low carbon steel, Scr. Metall., 12 (1978) 197–200.

[71] C. Fressengeas, Mechanics of dislocation fields, Wiley-ISTE, US, 2017.

[72] H. Jiang, Q. Zhang, X. Chen, Z. Chen, Z. Jiang, X. Wu, J. Fan, Three types of Portevin-Le Chatelier effects: experiment and modelling, Acta Mater. 55 (2007) 2219–2228.

[73] C. Bernard, J. Coër, H. Laurent, P.Y. Manach, M. Oliveira, L.F. Menezes, Influence of Portevin-Le Chatelier Effect on Shear Strain Path Reversal in an Al-Mg Alloy at Room and High Temperatures, Exp. Mech. 57 (2017) 405–415.

[74] P. G. McCormick, S. Venkadesan, C.P. Ling, Propagative instabilities: an experimental view, Scripta Metall. Mater. 29 (1993) 1159–1164.

[75] L.B. Zuev, S.A. Barannikova, Experimental study of plastic flow macro-scale localization process: Pattern, propagation rate, dispersion, Int. J. Mech. Sci. 88 (2014) 1–7.

[76] T.A. Lebedkina, Y. Bougherira, D. Entemeyer, M.A. Lebyodkin, I.V. Shashkov, Crossover in the scale-free statistics of acoustic emission associated with the Portevin-Le Chatelier instability, Scripta Mater. 148 (2018) 47–50.